\documentclass[journal]{IEEEtran}

\ifCLASSINFOpdf
\else
  \usepackage[dvips]{graphicx}
\fi
\usepackage{url}

\usepackage{color}
\usepackage{graphicx}
\usepackage{amsfonts}
\usepackage{amssymb}
\usepackage{stfloats}
\usepackage{cite}
\usepackage{psfrag}
\usepackage{subfigure}
\usepackage{amsmath}
\usepackage{array}
\usepackage{algorithm}
\usepackage{algpseudocode}
\usepackage{setspace}
\usepackage[english]{babel}
\usepackage{blindtext}
\usepackage{url}
\usepackage{verbatim}
\usepackage{color}
\usepackage{amsmath}
\usepackage{bm}
\usepackage{multirow}
\usepackage{booktabs}
\usepackage{makecell}
\usepackage{lineno}

\begin{document}

\title{High Accurate Time-of-Arrival Estimation with Fine-Grained Feature Generation for Internet-of-Things Applications}

\author{Guangjin Pan, Tao Wang, Shunqing Zhang, \IEEEmembership{Senior Member, IEEE}, and Shugong Xu, \IEEEmembership{Fellow, IEEE}
\thanks{This work was supported in part by the National Natural Science Foundation of China (NSFC) Grants under No. 61701293 and No. 61871262, the National Science and Technology Major Project (Grant No. 2018ZX03001009), the National Key Research and Development Program of China (Grant No. 2017YFE0121400), the Huawei Innovation Research Program (HIRP), and research funds from Shanghai Institute for Advanced Communication and Data Science (SICS).}
\thanks{G. Pan, T. Wang, S. Zhang, and S. Xu are with Shanghai Institute for Advanced Communication and Data Science, Key laboratory of Specialty Fiber Optics and Optical Access Networks, School of Information and Communication Engineering, Shanghai University, Shanghai, 200444, China (e-mail: \{guangjin\_pan, twang, shunqing, shugong\}@shu.edu.cn).}
\thanks{Corresponding Author: Shunqing Zhang.}
}

\markboth{IEEE Wireless Communications Letters, Vol. X, No. X, March 2020}
{Shell \MakeLowercase{\textit{et al.}}: Bare Demo of IEEEtran.cls for IEEE Journals}
\maketitle

\begin{abstract}
Conventional schemes often require extra reference signals or more complicated algorithms to improve the time-of-arrival (TOA) estimation accuracy. However, in this letter, we propose to generate fine-grained features from the full band and resource block (RB) based reference signals, and calculate the cross-correlations accordingly to improve the observation resolution as well as the TOA estimation results. Using the spectrogram-like cross-correlation feature map, we apply the machine learning technology with decoupled feature extraction and fitting to understand the variations in the time and frequency domains and project the features directly into TOA results. Through numerical examples, we show that the proposed high accurate TOA estimation with fine-grained feature generation can achieve at least 51\% root mean square error (RMSE) improvement in the static propagation environments and 38 ns median TOA estimation errors for multipath fading environments, which is equivalently 36\% and 25\% improvement if compared with the existing MUSIC and ESPRIT algorithms, respectively.
\end{abstract}


\IEEEpeerreviewmaketitle

\section{Introduction}

\IEEEPARstart{V}{ARIOUS} localization applications, together with the increasing number of smart terminals, trigger the rapid progress in the positioning technologies. Traditional global navigation satellite system (GNSS) and inertial navigation system (INS) are able to provide a continuous localization accuracy up to sub-meter level or even centimeter level with sufficient power supply. The recently proposed schemes, including enhanced Cell ID (E-CID), direction of arrival (DOA), observed time difference of arrival (OTDOA) \cite{NB4}, and fingerprint based solution \cite{mmwave_fingerprint}, provide alternative approaches for Internet-of-Things (IoT) devices served by cellular communication systems.
 
Among the above cellular signal based schemes, OTDOA only requires a single antenna observation with a reasonable localization accuracy, which is of great interest for low power IoT applications\footnote{{Although using wideband signals is quite efficient to improve the estimation accuracy, it contradicts with the design principle of low power IoT systems.}}, especially in the recent deployed narrowband IoT (NB-IoT) systems \cite{NB4}. A typical issue is the {low} sampling rate with 1.92 Mbps, which corresponds to the observation of 520 ns in the time domain, and the resolution of time-of-arrival (TOA) estimation accuracy is around hundred meters \cite{SAGE}. To eliminate this effect, estimation of signal parameters via rotational invariance techniques (ESPRIT) \cite{ESPRIT1}, multiple signal classification (MUSIC) \cite{MUSIC}, as well as the maximum peak-to-leaking ratio (MPLR) \cite{MPLR} algorithms have been proposed to improve the TOA estimation accuracy up to fractional observation periods. Upsampling technique has also been proposed in \cite{Iot-resolution} to improve the resolution of cross-correlations and achieves 40\% enhancement of TOA estimation accuracy. Another approach adopts deep convolutional neural network (DCNN)\cite{DCNN} to extract the intrinsic features, and the achievable accuracy is around two meters in WiFi system.

Although the aforementioned schemes can improve the TOA estimation performance at the expense of complexity, there are still many practical challenges in the low power IoT scenarios. For example, ESPRIT and MUSIC algorithms require real-time fast Fourier transform (FFT) and channel frequency response (CFR) estimation before the subspace manipulation, which becomes a heavy burden if the TOA estimation needs to be performed frequently. For the DCNN based solution, the adaptation capability to fit different channel environments is usually limited, since it highly relies on the training data to obtain satisfied estimation results \cite{DCNN}. Moreover, the MPLR algorithm can achieve reasonable estimation accuracy only if the reference signals are uniformly distributed. 

In this letter, we propose a high accurate TOA estimation scheme with fine-grained feature generations for IoT applications. Specifically, we generate both the full band and resource block (RB) based reference signals, and calculate the cross-correlations with sampled observations to generate the feature map. Using this spectrogram-like cross-correlation feature map, we apply the machine learning technology with decoupled feature extraction and fitting to understand the variations in the time and frequency domains. In this way, we project the features directly into TOA results. Numerical results show that the proposed high accurate TOA estimation method can achieve more than 51\% and 25\% RMSE improvement in the static and multi-path fading environments, respectively.

\section{Backgrounds} \label{sec2}
\begin{figure*}[tb]
\centering 
\includegraphics[height=1.8in,width=6.7in]{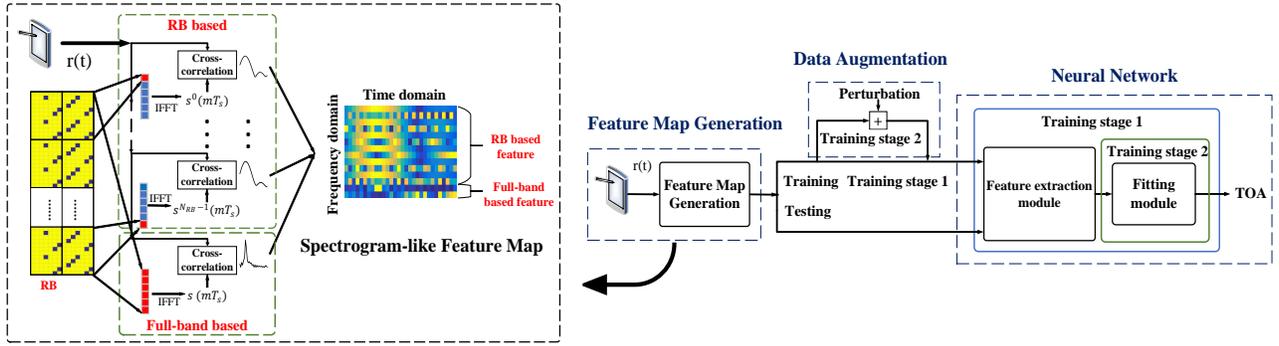} 
\caption{Schematic diagram of the proposed TOA estimation scheme. The input of the neural network is the feature map constructed by the amplitude and phase of the full band and RB based cross-correlation function. After training the network using two-stage training method, the system selects a fitting module to estimate the TOA according to the channel statistical information.}
\label{TOA_estimation} 
\end{figure*}
Consider a general IoT system using orthogonal frequency division multiplexing (OFDM) scheme with $K$ subcarriers. Denote $d^{k}_{p}$ to be position reference signal (PRS) of baseband for the $k^{th}$ subcarrier and the $p^{th}$ time slot, and the corresponding time domain representation, $s_p\left(t\right)$, is simply given by, $s_p\left(t\right) = \sum_{k=0}^{K-1} d^{k}_{p}e^{j2\pi k\triangle{f}\left(t-N_{cp}T_s\right)}$, for $t \in \left[0, (K+N_{cp}-1)T_s \right]$. $N_{cp}$ and $T_s$ denote the length of cyclic prefix (CP) and the sampling period, respectively, and $\triangle{f}$ is the subcarrier spacing. Denote $\tau_l$ to be the propagation delay of the $l^{th}$ fading path, the received signal at the baseband can be expressed as,
\begin{eqnarray}\label{equ2-2}
r_p\left(t\right)=\sum_{l=0}^{L-1}\ h(\tau_l) s_p\left(t-\tau_l\right) + w\left(t\right),
\end{eqnarray}
where {$h(\tau_l)$} is the channel response corresponding of the $l^{th}$ fading path, and $w\left(t\right)$ denotes the additive white Gaussian noise (AWGN) with zero mean and variance $\sigma^2$. At the receiver side, by multiplying the conjugate of transmitted signal $s_p(t)$, the cross-correlation $R(\cdot)$ can be calculated by, 
\begin{eqnarray}
R(mT_s) & = & \sum_{n=m}^{N_t+m-1} r_{p}\left(nT_s\right) {s}_{p}^{\ast}\left((n-m)T_s\right) \nonumber \\
& = & \sum_{l=0}^{L-1} h(\tau_l) \gamma\left(mT_s-\tau_{l}\right) +\hat{w}\left(mT_s\right),
\end{eqnarray}
where $\gamma\left(mT_s-\tau_{l}\right)=\sum_{n=m}^{N_t+m-1}\ s_{p}\left(nT_s\right) {s}_{p}^{\ast}((n-m)T_s-$ $\tau_l)$. $N_t$ denotes the size of searching window, and $\hat{w}\left(mT_s\right)$ is the remaining noise term.

Theoretically speaking, an exhaustive based scheme to search over all the possible $L$, $\{\tau_l\}$, and {$\{h(\tau_l)\}$} is usually required to minimize the localization error\cite{Multipath_delay}. However, to reduce the search dimension in the practical implementation, we often fix the number of $L$ by observing the signal characteristics in the cross-correlation function $R(mT_s)$ \cite{Iot-resolution} or some pre-defined numbers \cite{SAGE}. Thus, the optimal delay profiles as well as channel responses can be estimated via\footnote{For illustration purpose, we consider one dimensional received sequences and the extension to two dimensional case is straight forward. In addition, we allow the fractional sampling period and generate {$h(\tau_l)$} according to \cite{MUSIC}.},

\begin{small}
\begin{eqnarray}
\label{equ2-3}
\{\tau_{l}^{\star}\},\{h^{\star}(\tau_l)\} = \mathop{\arg \min}_{\{\tau_l\},\{h(\tau_l)\}} \sum_{m=0}^{M-1}\big| R(mT_s)-\sum_{l=0}^{L-1} h(\tau_l) \gamma(mT_s-\tau_{l})\big|^2,
\end{eqnarray}
\end{small}where $M$ indicates the observation window length of the cross-correlation function $R(mT_s)$. By taking the minimum value of the delay profile $\{\tau_l\}$, we obtain the estimated TOA through $\tau^{\star} = \min_{l} \{\tau_{l}^{\star}\}$.

\section{Proposed TOA estimation scheme}\label{sec3}

In general, the above optimization problem for TOA estimation requires a joint search of unknown number of fading paths, channel coefficients, and the possible TOAs. Conventional estimation algorithms, such as SAGE \cite{SAGE}, are able to obtain rough results via iterative methods, while the estimation accuracy is difficult to improve, especially when the sampling rate is limited. In this section, we propose a deep learning based TOA estimation scheme together with several augmentation techniques to solve this issue.

\subsection{Overview of the Proposed Scheme}\label{sec3-1}

The general schematic diagram of the proposed TOA estimation scheme is shown in Fig.~\ref{TOA_estimation}, where two main improvements are elaborated below. 

\subsubsection{Fine-grained Feature Generation}
In order to improve the resolution of cross-correlation observations, we generate the reference signals in RB based manner\footnote{According to the 3GPP specification \cite{TS36.211}, one resource block contains 12 subcarriers and 1 time slot with 7 OFDM symbols.} at the receiver side. The time domain representation is given by, $s_p^v(t) = \sum_{k \in \Omega^{v}} d^{k}_{p}e^{j2\pi k\triangle{f}\left(t-N_{cp}T_s\right)}$, where $\Omega^v$ denotes the set of subcarrier indices in the $v^{th}$ RB and the cardinality of $\cup_v \Omega^{v}$ equals to $K$. Following the same procedures as introduced in Section~\ref{sec2}, we have the RB based cross-correlations, $\{R^v(mT_s)\}$, as follows. 
\begin{eqnarray}
R^v(mT_s) = \sum_{l=0}^{L-1} h(\tau_l) \gamma^v\left(mT_s-\tau_{l}\right) +\hat{w^v}\left(mT_s\right),
\end{eqnarray}
where $\gamma^v\left(mT_s-\tau_{l}\right) = \sum_{n=m}^{N_t+m-1}\ s_{p}\left(nT_s\right) {s}_{p}^{v,\ast}((n-m)T_s-$ $\tau_l)$. With the above manipulations, the input spaces of the proposed TOA estimation scheme can incorporate both the full band cross-correlation $R(mT_S)$ and the RB based cross-correlations $\{R^v(mT_s)\}$. Compared with the conventional approach as defined in \eqref{equ2-3}, it has more fine-grained cross-correlation observations and greatly improves the resolution of cross-correlation based feature maps as shown in Fig.~\ref{TOA_estimation}. 

\subsubsection{Decoupled Feature Extraction and Fitting}
The straight forward method for TOA estimation using the above fine-grained features is to compute TOAs based on the collected $R(mT_S)$ and $R^v(mT_s)$ for $m \in [0, M-1]$ { using (\ref{equ2-3}) and ensemble results through majority voting or simple averaging.} However, this type of scheme fails to explore the interrelationship among different cross-correlation observations. Therefore, a more reasonable framework is to directly minimize the mean square errors (MSE) of TOA estimation through,
\begin{eqnarray}
\mathop{\textrm{minimize}}_{\mathcal{F}(\cdot)} && \big|\mathcal{F} \big(R(mT_S),\{R^v(mT_S)\}, \Sigma_{\{h(\tau_l)\}, \{\tau_l\}} \big) - \tau \big|^2, \nonumber \\
\textrm{subject to} && \tau = \min_l \{\tau_l\}, \label{eqn:origin}
\end{eqnarray}
where $\Sigma_{\{h(\tau_l), \tau_l\}}$ denotes the available statistical information of channel fading coefficients and propagation delays, {and $\mathcal{F}(\cdot)$ represents the mapping function from the obtained cross-correlations and the historical information to the estimating TOA result.} A straight forward idea for easy deployment is to decouple the environment dependent parameters $\Sigma_{\{h(\tau_l), \tau_l\}}$ and the observation dependent parameters $R(mT_S),\{R^v(mT_s)\}$. Inspired by \cite{NIPS}, we construct two concatenated representative functions $\mathcal{H}_e$ and $\mathcal{H}_f$ with auxiliary parameters $\theta_{e}$ and $\theta_{f}$ to approximate the original function $\mathcal{F}(\cdot)$. { $\mathcal{H}_e$ extracts general features of TOA from the cross-correlation functions of the received signal, and $\mathcal{H}_f$ uses the features extracted by $\mathcal{H}_e$ to fit the TOA estimation result. Applying this notation, the} original minimization becomes,
\begin{eqnarray}
\mathop{\textrm{minimize}}_{\theta_{e},\theta_{f}} && \big|\mathcal{H}_f \big( \mathcal{H}_e \left(R(mT_S),\{R^v(mT_s)\};\theta_{e})\right), \nonumber \\ 
&& \Sigma_{\{h(\tau_l)\}, \{\tau_l\}};\theta_{f} \big) - \tau \big|^2, \nonumber \\
\textrm{subject to} && \tau = \min_l \{\tau_l\}.
\end{eqnarray}
Through the above scheme, we can isolate the deployment of the feature extraction process $\mathcal{H}_e(\cdot;\theta_e)$ and the TOA fitting process $\mathcal{H}_f(\cdot;\theta_f)$ as depicted in Fig.~\ref{TOA_estimation}, and the corresponding advantages are listed below. {First, we can adopt a common implementation strategy for IoT sensors with shared feature extraction. After obtaining the extracted features, a customized TOA estimation module at the central cloud or mobile edge units is applied to analyse them using environment dependent models. Through this approach, it can perfectly solve the aforementioned issues.} Second, the data security and communication issues can be partly solved using the extracted feature patterns rather than the original cross-correlation observations. Last but not least, the adaptation to new wireless environments is much easier, since only the fitting modules need to be updated in order to achieve satisfactory performance.

\subsection{Neural Network Design}

\begin{table}[ht]\footnotesize 
	\centering 
	\caption{Detailed Neural Network Configuration} 
	\label{table1}
  \renewcommand\arraystretch{1.5}
	\begin{tabular}{|c|c|c|c|} 
    \bottomrule
		Functions&Layers&Output Dimension&Activation \\ 
		\hline \bottomrule
  	\multirow{4}{1.5cm}{\centering Feature extraction $\mathcal{H}_e(\cdot;\theta_e)$}&Input& M*(1+$N_{RB}$)*2& ReLU \\ \cline{2-4}
		&Conv2D 1& $\frac{M}{2}*N_{RB}*2$ & ReLU\\ \cline{2-4}
		&Conv2D 2& $\frac{M}{4}*\frac{N_{RB}}{2}*2$ & ReLU\\ \cline{2-4}
		&Flatting& $\frac{MN_{RB}}{4}$ & 
		\\ 
		\bottomrule
		\multirow{5}{1.5cm}{\centering Fitting $\mathcal{H}_f(\cdot;\theta_f)$}&Input& $\frac{MN_{RB}}{4}$&ReLU\\ \cline{2-4}
		&FC 1& 32 & ReLU\\ \cline{2-4}
		&FC 2& 8 & ReLU\\ \cline{2-4}
		&FC 3 (output)& 1 & Linear\\ \bottomrule		
	\end{tabular}
\end{table}

Different from the conventional approach to predict the entire propagation delay profile before TOA estimation, we directly model the functions $\mathcal{H}_e(\cdot;\theta_e)$ and $\mathcal{H}_f(\cdot;\theta_f)$ through neural networks. The corresponding network configuration and loss function are elaborated below. 
\subsubsection{Network Configuration}\label{Network_configutation}
The detailed network configuration contains two-dimensional convolutions (Conv2D) and fully connected (FC) layers with Rectified Linear Unit (ReLU) activation, as shown in Table \ref{table1}. Meanwhile, in order to meet different input dimensions, we use a parameterized output dimension with respect to the total observation window length, $M$, and the total number of RB, $N_{RB}$, rather than some specific numbers.

\subsubsection{Loss Function}


{Due to the limited sampling rate, the resolution of TOA estimation is limited. Meanwhile, a more fine-grained classification will increase the dimension of output vector, which increases the neural network complexity as well as the computational overhead.} To address this issue, we directly adopt the regression model to predict the TOA result $\tau$ and the associated loss function, $\mathcal{L}_{\tau}$, is chosen to be the MSE between predicted results and true TOA values\footnote{{The true TOA values are obtained by directly dividing the Euclidean distances between transmitters and receivers over the speed of light.}},
\begin{eqnarray}
\mathcal{L}_\tau & = & \big|{\mathcal{H}_f} \big( \mathcal{H}_e \left(R(mT_S),\{R^v(mT_s)\};\theta_{e})\right), \nonumber \\ 
&& \Sigma_{\{h(\tau_l)\}, \{\tau_l\}};{\theta_{f} }\big) - \tau \big|^2. \end{eqnarray}

\subsection{Training Skills}
In order to train two concatenated blocks in the proposed neural network architecture, we propose a two-stage training scheme with data augmentation technique to reduce the training data requirement. 
\subsubsection{Two-stage Training}
In the first stage, we simultaneously train two concatenated blocks using the data sets generated under different channel fading environments, which enables the feature extraction module to learn sufficient TOA-related features. In the second stage, we freeze the feature extraction part to be $\hat{\mathcal{H}}_e(\cdot;\hat{\theta}_e)$, and fine-tune the fitting module with low cost training samples. It is worth to mention that the training data in the second stage can be collected from remote entities with feature extraction capabilities or generated by the data augmentation scheme as explained later.

\subsubsection{Perturbation based Data Augmentation}
Denote $\delta(mT_S)$ and $\{\delta^v(mT_S)\}$ to be the random perturbation on the full band and RB based cross-correlation observations, respectively. The corresponding output vectors of the feature extraction module, $\eta_{\delta}$, are given by $\eta_{\delta} = \hat{\mathcal{H}}_e(R(mT_S)+\delta(mT_S), \{R^{v}(mT_S) + \delta^v(mT_S)\};\hat{\theta}_e)$. By pairing $\eta_{\delta}$ with the original TOA value $\tau$, we can enlarge the training data set for the second stage optimization, which is defined as,
\begin{eqnarray}
\mathop{\textrm{minimize}}_{\theta_{f}}&& \big|\mathcal{H}_f \big( \eta_{\delta}, \Sigma_{\{h(\tau_l)\}, \{\tau_l\}};\theta_{f} \big) - \tau \big|^2.
\end{eqnarray}
It is worth to mention that we generate\footnote{{According to \cite{Perturbation2}, the perturbation can be utilized to make neural networks smoother and more robust without significantly sacrificing the system performance. In a real-life implementation, we also need to filter out abnormal data to eliminate outliers as proposed in \cite{outlier}.}} $\delta(mT_S)$ and $\{\delta^v(mT_S)\}$ according to zero mean Gaussian distribution with the variations much less than the absolute values of $R(mT_S)$ and $\{R^{v}(mT_S)\}$.

\section{Numerical Results}
In this section, we provide some numerical results to demonstrate the advantages of proposed TOA estimation scheme by comparing with state-of-the-art results\footnote{{The data set as well as the source codes in this letter are available at {https://github.com/XLPolar/TOA\_estimation}}.}. In particular, we consider a typical NB-IoT scenario with sampling rate $1.92$ Mbps, which is equivalent to $T_s = 520$ ns and $N_{RB} = 6$. The numerical simulations are performed under three channel conditions, e.g., static propagation environments (case 1), multipath fading environments including Extended Pedestrian A model with a doppler frequency of 5Hz (EPA-5Hz) (case 2) and Extended Vehicular A model with a doppler frequency of 5Hz (EVA-5Hz) (case 3) \cite{TS36.104}. Without the prior statistical knowledge of fading environments, we choose\footnote{{In fact, our proposed algorithm can be extended to other OFDM-based IoT communication systems, such as WiFi or LTE-M, since we only choose extremely low sampling rate, limited RB numbers, and no exact channel statistical knowledge in our simulation settings.}} $\Sigma_{\{h(\tau_l)\}, \{\tau_l\}}$ to be $\{\textrm{case 1}, \textrm{case 2}, \textrm{case 3}\}$, and the mapping scheme of PRS follows 3GPP Release 15 standard \cite{TS36.211}. 

\subsection{Effects of Training Skills}

\begin{figure}[tb]
\centering 
\includegraphics[height=1.8in,width=2.5in]{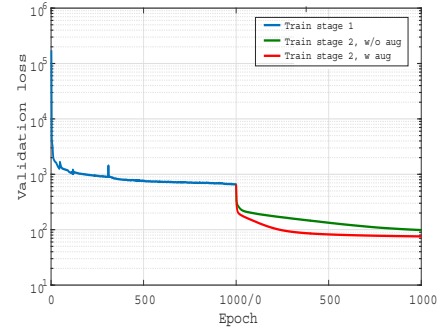} 
\caption{Validation loss comparison using different training skills.}
\label{loss} 
\end{figure}

In the first experiment, we compare the validation losses and cumulative distribution function (CDF) of TOA estimation errors using different training skills to demonstrate the advantages of the proposed scheme. As illustrated before, we accumulate the accurate TOA results under three cases to train the proposed {neural networks} and test them under training strategies. {In Fig.~\ref{loss}, we share the training result in terms of the validation loss in the first stage (blue line), and compare the validation losses of the proposed scheme with (red line) and without (green line) data augmentation strategies. Each epoch represents one training process using all training data. As shown in this figure, the validation loss of stage 1 converges to $10^{4}$ and that of stage 2 can achieve a better result, $10^{3}$. If we can apply perturbation based data augmentation, we can obtain another 20\% reduction, e.g., from $10^{3}$ to $8 \times 10^{2}$.}

\begin{figure}[tb]
\centering 
\includegraphics[height=1.8in,width=2.5in]{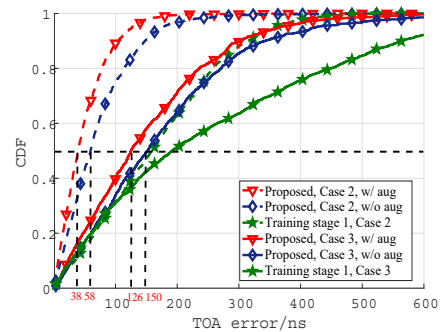} 
\caption{TOA estimation performance of proposed training skills. The red, blue, and green lines represent the proposed algorithm with/without data augmentation, and the intermediate results after the first training stage, respectively.}
\label{skill} 
\end{figure}

Fig.~\ref{skill} compares the CDF of TOA estimation errors for different training strategies, e.g., the effects of {two-stage training scheme} and perturbation based data augmentation under the multipath fading environments. As shown in this figure, the proposed two-stage training scheme can achieve at least 90 ns and 40 ns TOA error reduction for case 2 and case 3, respectively, if compared with one-stage training. 
The numerical results show that the above approach can provide another 20 ns improvement, which corresponds to 114 ns and 70 ns TOA error reduction for case 2 and case 3 in total.

\subsection{Static Propagation Environments}

\begin{figure}[thb]
\centering 
\includegraphics[height=1.8in,width=2.5in]{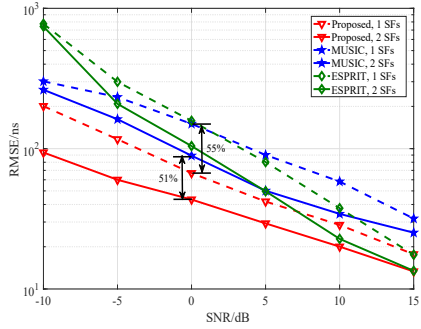} 
\caption{Performance comparison with baseline algorithms. When the duration of the PRS is 1 or 2 subframes, the proposed algorithm performs better than the baseline algorithms. }
\label{fig1} 
\end{figure}

Fig.~\ref{fig1} shows the RMSE versus signal-to-noise ratio (SNR) performance of different TOA estimation schemes. PRSs within one and two subframes are utilized to test different schemes. With the proposed fine-grained feature generation and decoupled feature extraction, we can achieve a better RMSE performance, especially in the low SNR regime As an example, when SNR equals to $0$ dB, the proposed scheme can achieve more than 51\% RMSE improvement if PRSs within one or two subframe durations are utilized, respectively. Since ESPRIT relies on the phase differences of each channel, it can obtain similar RMSE performance in the static propagation environment with high SNR value as observed from Fig.~\ref{fig1}. However, in the practical multipath fading environment, especially for low power IoT applications, the obtained SNR is usually limited and our proposed algorithm is more reliable. In this experiment, we apply the same feature extraction module to show that the decoupled feature extraction and fitting strategy are feasible for different PRSs' configurations.

\begin{figure}[tbp]
\centering 
\subfigure[CDF of TOA estimation error comparison in case 2.] 
{
\begin{minipage}{1\linewidth}
\centering
\includegraphics[height=1.8in,width=2.5in]{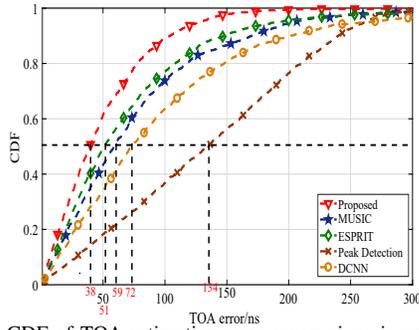} 
\end{minipage}
}
\subfigure[CDF of TOA estimation error comparison in case 3.] 
{
\begin{minipage}{1\linewidth}
\centering
\includegraphics[height=1.8in,width=2.5in]{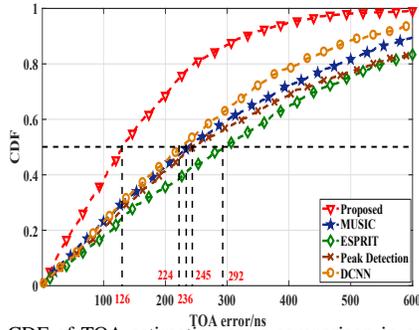} 
\end{minipage}
}

\caption{CDF of TOA estimation error in multipath channels. Red line represents the proposed algorithm and the remaining lines represent four baseline schemes, including MUSIC, ESPRIT, peak detection \cite{SAGE} and DCNN.} 
\label{EPA1} 
\end{figure}

\subsection{Multipath Fading Environments}

Fig.~\ref{EPA1} compares the CDF of TOA estimation errors in the multipath fading environments, where PRSs with two subframe durations are configured. The proposed TOA estimation scheme with fine-grained feature generation can achieve the median errors of 38 ns and 126 ns for case 2 and case 3, respectively. In other words, the proposed TOA estimation scheme can achieve 36\% and 47\% improvement if compared with the MUSIC algorithm, 25\% and 57\% improvement if compared with the ESPRIT algorithm, or 47\% and 44\% improvement if compared with the DCNN scheme. In the above numerical simulations, we also plot the estimation errors of the conventional scheme as defined by \eqref{equ2-3} for reference purpose. {Meanwhile, we list the running time of different algorithms in Table \ref{table2} for the sake of completeness.} Based on the above results, we conclude that the proposed method can outperform the existing baseline schemes and achieve a significant TOA estimation accuracy improvement for static propagation as well as multipath fading environments. { Meanwhile, the computational complexity of proposed scheme is also lower than other baseline algorithms.}

\begin{table}[tbp]\footnotesize 
	\centering 
	\caption{RUNNING TIME COMPARISON FOR DIFFERENT ALGORITHMS} 
	\label{table2}
  \renewcommand\arraystretch{1.5}
	\begin{tabular}{|c|c|c|c|} 
    \bottomrule
		Algorithm&Running Time&Algorithm&Training/Test Time\\
		\hline
		MUSIC&457.4 s&Proposed&1792 s / 1.8 s\\ 
		\hline
  	ESPRIT&6.1 s&DCNN & 5423 s / 2.9 s \\ \hline
	\end{tabular}
\end{table}

\section{Conclusion}

In this paper, we propose a high accurate ToA estimation algorithm based on fine-grained feature generation using a spectrogram-like cross-correlation feature map. The idea of decoupled feature extraction and fitting is then applied to improve the estimation accuracy and system robustness. A two-stage training strategy with perturbation based data augmentation scheme is utilized to fine-tune the feature extraction and fitting modules. {Through some numerical results, we believe the proposed estimation scheme is promising to achieve more accurate TOA results than conventional methods.}



\begin{thebibliography}{10}
\providecommand{\url}[1]{#1}
\csname url@samestyle\endcsname
\providecommand{\newblock}{\relax}
\providecommand{\bibinfo}[2]{#2}
\providecommand{\BIBentrySTDinterwordspacing}{\spaceskip=0pt\relax}
\providecommand{\BIBentryALTinterwordstretchfactor}{4}
\providecommand{\BIBentryALTinterwordspacing}{\spaceskip=\fontdimen2\font plus
\BIBentryALTinterwordstretchfactor\fontdimen3\font minus
  \fontdimen4\font\relax}
\providecommand{\BIBforeignlanguage}[2]{{%
\expandafter\ifx\csname l@#1\endcsname\relax
\typeout{** WARNING: IEEEtran.bst: No hyphenation pattern has been}%
\typeout{** loaded for the language `#1'. Using the pattern for}%
\typeout{** the default language instead.}%
\else
\language=\csname l@#1\endcsname
\fi
#2}}
\providecommand{\BIBdecl}{\relax}
\BIBdecl

\bibitem{NB4}
{X. Lin, J. Bergman, et al.}, ``Positioning for the internet of things: A 3gpp
  perspective,'' \emph{IEEE Commun. Mag.}, vol.~55, no.~12, pp. 179--185, Dec
  2017.

\bibitem{mmwave_fingerprint}
{J. Gante, G. Falcão and L. Sousa}, ``Deep learning architectures for accurate
  millimeter wave positioning in 5g,'' \emph{Neural Process Lett}, vol.~47,
  no.~1, pp. 487--514, Aug. 2020.

\bibitem{SAGE}
S.~{Hu}, X.~{Li}, and F.~{Rusek}, ``On time-of-arrival estimation in {NB-IoT}
  systems,'' in \emph{Proc. IEEE WCNC'19}, Apr. 2019, pp. 1--6.

\bibitem{ESPRIT1}
R.~{Roy} and T.~{Kailath}, ``Esprit-estimation of signal parameters via
  rotational invariance techniques,'' \emph{IEEE Trans. Acousr., Speech, Signal
  Processing}, vol.~37, no.~7, pp. 984--995, July 1989.

\bibitem{MUSIC}
X.~{Li} and K.~{Pahlavan}, ``Super-resolution toa estimation with diversity for
  indoor geolocation,'' \emph{IEEE Trans. Wireless Commun.}, vol.~3, no.~1, pp.
  224--234, Jan 2004.

\bibitem{MPLR}
Z.~{He}, Y.~{Ma}, and R.~{Tafazolli}, ``Improved high resolution toa estimation
  for ofdm-wlan based indoor ranging,'' \emph{IEEE Wireless Commun. Lett.},
  vol.~2, no.~2, pp. 163--166, Jan. 2013.

\bibitem{Iot-resolution}
S.~Hu, A.~Berg, X.~Li, and F.~Rusek, ``{Improving the Performance of OTDOA
  Based Positioning in NB-IoT Systems },'' in \emph{Proc. GLOBECOM'17}, Jan.
  2017, pp. 1--7.

\bibitem{DCNN}
O.~{Bialer}, N.~{Garnett}, and D.~{Levi}, ``A deep neural network approach for
  time-of-arrival estimation in multipath channels,'' in \emph{Proc. IEEE
  ICASSP'18}, Apr. 2018, pp. 2936--2940.

\bibitem{Multipath_delay}
{J. J. Fuchs}, ``Multipath time-delay detection and estimation,'' \emph{IEEE
  Trans. Signal Process.}, vol.~47, no.~1, pp. 237--243, Jan. 1999.

\bibitem{TS36.211}
{3GPP TS 36.211}, ``Technical specification group radio access network; evolved
  universal terrestrial radio access {(E-UTRA)}; physical channels and
  modulation,'' {}V15.7.0, May. 2019.

\bibitem{NIPS}
{J. Yosinski, J. Clune, et al.}, ``How transferable are features in deep neural
  networks?'' in \emph{Proc. NIPS'14}, Dec. 2014, pp. 3320--3328.

\bibitem{Perturbation2}
{C. Xiang, S. Zhang, et al.}, ``Robust sub-meter level indoor localization with
  a single wifi access point—regression versus classification,'' \emph{IEEE
  Access}, vol.~7, pp. 146\,309--146\,321, Oct. 2019.

\bibitem{outlier}
H.~{Wang}, M.~J. {Bah}, and M.~{Hammad}, ``Progress in outlier detection
  techniques: A survey,'' \emph{IEEE Access}, vol.~7, pp. 107\,964--108\,000,
  Aug. 2019.

\bibitem{TS36.104}
{3GPP TS 36.104}, ``Technical specification group radio access network; evolved
  universal terrestrial radio access {(E-UTRA)}; base station {(BS)} radio
  transmission and reception,'' {V16.4.0, Dec. 2019}.

\end{thebibliography}
\end{document}